\documentclass[%
superscriptaddress,
 amsmath,amssymb,
 aps, 
twocolumn
]{revtex4-2}

\usepackage{graphicx}
\usepackage{dcolumn}
\usepackage{bm}
\usepackage{xcolor}
\begin{document}

\preprint{APS/123-QED}

\title{\textbf{Engineering correlated disorder for tailored light scattering} 
}%

\author{Denis Langevin}
 \email{Contact author: denis.langevin@polytechnique.org}
\affiliation{%
Aix Marseille Univ, CNRS, Université de Toulon, IM2NP, UMR 7334, F-13397 Marseille, France 
}%
\author{Emma Bosbaty-Galliot}%
\affiliation{%
 Universit\'e Clermont Auvergne, CNRS, Institut Pascal, F-63000 Clermont-Ferrand, France
}
\author{Emmanuel Centeno}%
\affiliation{%
 Universit\'e Clermont Auvergne, CNRS, Institut Pascal, F-63000 Clermont-Ferrand, France
}
\author{Pauline Bennet}%
\affiliation{%
Aix Marseille Univ, CNRS, Université de Toulon, IM2NP, UMR 7334, F-13397 Marseille, France  
}
\author{Rémi Carminati}%
\affiliation{%
 Institut Langevin, ESPCI Paris, Université PSL, CNRS, F-75005 Paris, France
}
\affiliation{Institut d’Optique Graduate School, Université Paris-Saclay, Palaiseau, France}

\author{Bodo D. Wilts}
\affiliation{Department of Chemistry and Physics of Materials, University of Salzburg, AT-5020 Salzburg, Austria}

\author{Patrick Bouchon}%
\affiliation{%
 DOTA, ONERA, Université Paris-Saclay, Palaiseau, France
}
\author{Antoine Moreau}%
\affiliation{%
 Universit\'e Clermont Auvergne, CNRS, Institut Pascal, F-63000 Clermont-Ferrand, France
}%

\newcommand{\pauline}[1]{{\textcolor{violet}{#1}}}

\date{\today}

\begin{abstract}
Correlated disorder is known to shape light scattering in ways uncorrelated disorder cannot, from hyperuniform transparency to the structural colors of naturally occurring structures. What has been missing in photonics is a direct link between the disorder and the scattering pattern it produces. Here we show that adding correlated noise to a periodic array splits the scattering pattern
into three distinct components: diffraction peaks, a diffuse background, and correlation halos. Often mistaken for broadened diffraction peaks, these halos are in fact independent features: their positions are set by the correlation range, meaning that they can appear between Bragg peaks, and — crucially — they persist far beyond the regime where the diffraction peaks vanish. Shaping the disorder itself offers further control: tuning the noise distribution suppresses selected diffraction peaks, while tuning the correlation statistics moves the halos away from the Bragg positions. This approach reproduces the scattering signatures of natural photonic structures, such as Morpho butterfly wings, and reveals multiple pathways from order to disorder, each with distinct optical properties. It also offers a practical route to inverse design — finding the disorder that produces a desired scattering pattern. This establishes scattering as a designable quantity, expanding the toolkit for metasurfaces and structural colors.

\end{abstract}

\maketitle



The study of correlated disorder, and especially the discovery of hyperuniformity, marked a turning point in the understanding of disordered systems, demonstrating that correlations in disorder can lead to unexpected transparency and other remarkable properties \cite{torquato2003local, leseur2016high, bigourdan2019enhanced, vynck2023light, Klatt_phoamtonics, vynck2022visual}. This finding suggested that the transition from disorder to order could be controlled through correlation strength \cite{Yu2021engineered}. However, this perspective implicitly assumes only a single path: increasing correlations progressively transforms disorder into order. In reality, when studying quasi-periodic structures abundant in nature, the phase and property space of order and disorder appears far richer \cite{Jacucci_corrdis, StavengaKingfisher, DjeghdiAnt, YinMacaw, chen2025emergent}, leading to diverse diffraction patterns. Butterfly wing scales, bird feathers, and beetle cuticles, for instance, often exhibit quasi-periodic arrangements of nanostructures that produce distinctive optical signatures - broad scattering features centered on the positions where the Bragg peaks of the perfect crystal would lie \cite{vukusic1999quantified, schroeder2018s, mouchet2020optical, wilts2020ultra, moyroud2017disorder, rothammer2021tailored}.

However, when one attempts to model these structures by adding noise to periodic lattices, the standard approach using uncorrelated (white) noise completely destroys the diffraction peaks, leaving only diffuse scattering \cite{saito2011numerical, johansen2014optical, song2017reproducing}. This failure has sparked debates about whether the observed patterns truly arise from perturbed periodicity or require entirely different mechanisms. It has also made the optical properties of these natural structures difficult to reproduce accurately \cite{song2017reproducing, sterl2021shaping, herkert2023influence}.

A way out of this impasse comes, perhaps surprisingly, from the crystallography of imperfect crystals, where the effect of correlated positional disorder on scattering has long been described analytically \cite{welberry1985diffuse,leroy2004effects}. We use this framework, but with a different mindset. In crystallography, the disorder is a fixed property of a given sample and the aim is to infer it from the measured diffraction pattern. In optics, disorder can instead be designed at will \cite{lalanne2024disordered}, starting from the bounded perturbation of a periodic array — a disorder for which the Bragg peaks can survive. Seen this way, the framework of correlated disorder suggests that many possibilities remain unexplored. As a consequence, disorder shifts from a measurable property to a design tool.

Here, starting from the premise that in optics disorder can be controlled completely, we revisit this analytical framework and extend it to a fully tunable disorder. Adding controlled noise to a periodic array, we show that the scattering pattern keeps sharp Bragg peaks and a diffuse background — but, above all, contains broad features we call "correlation halos". These halos are remarkably robust to noise and persist even when the diffraction peaks have vanished; we show that they account for the angular distribution of light scattered by the wings of Morpho butterflies. We then explore the possibilities this design space opens, and show that one can suppress selected diffraction peaks, or place the correlation halos away from the Bragg positions — giving access to structures whose optical properties remain unexplored.

\begin{figure*}[ht]
    \centering
    \includegraphics[width=0.45\linewidth]{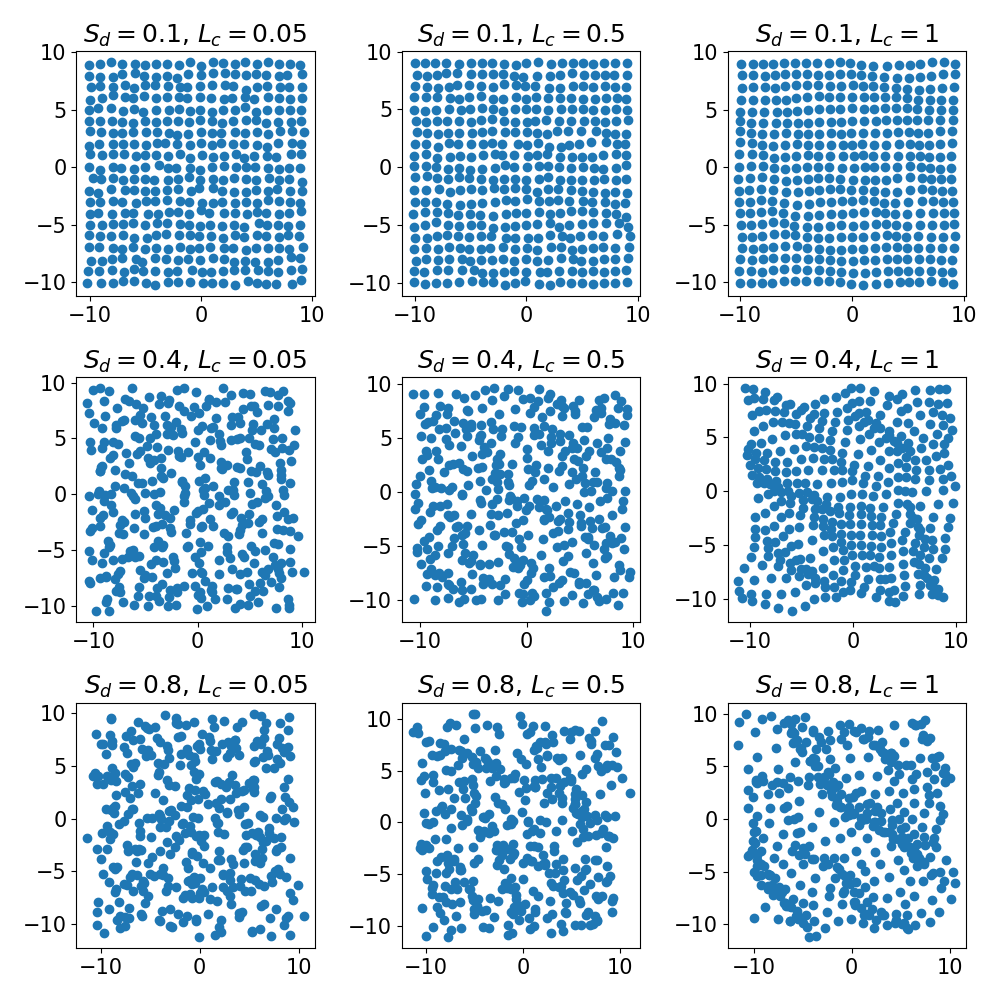}
    \quad
    \includegraphics[width=0.45\linewidth]{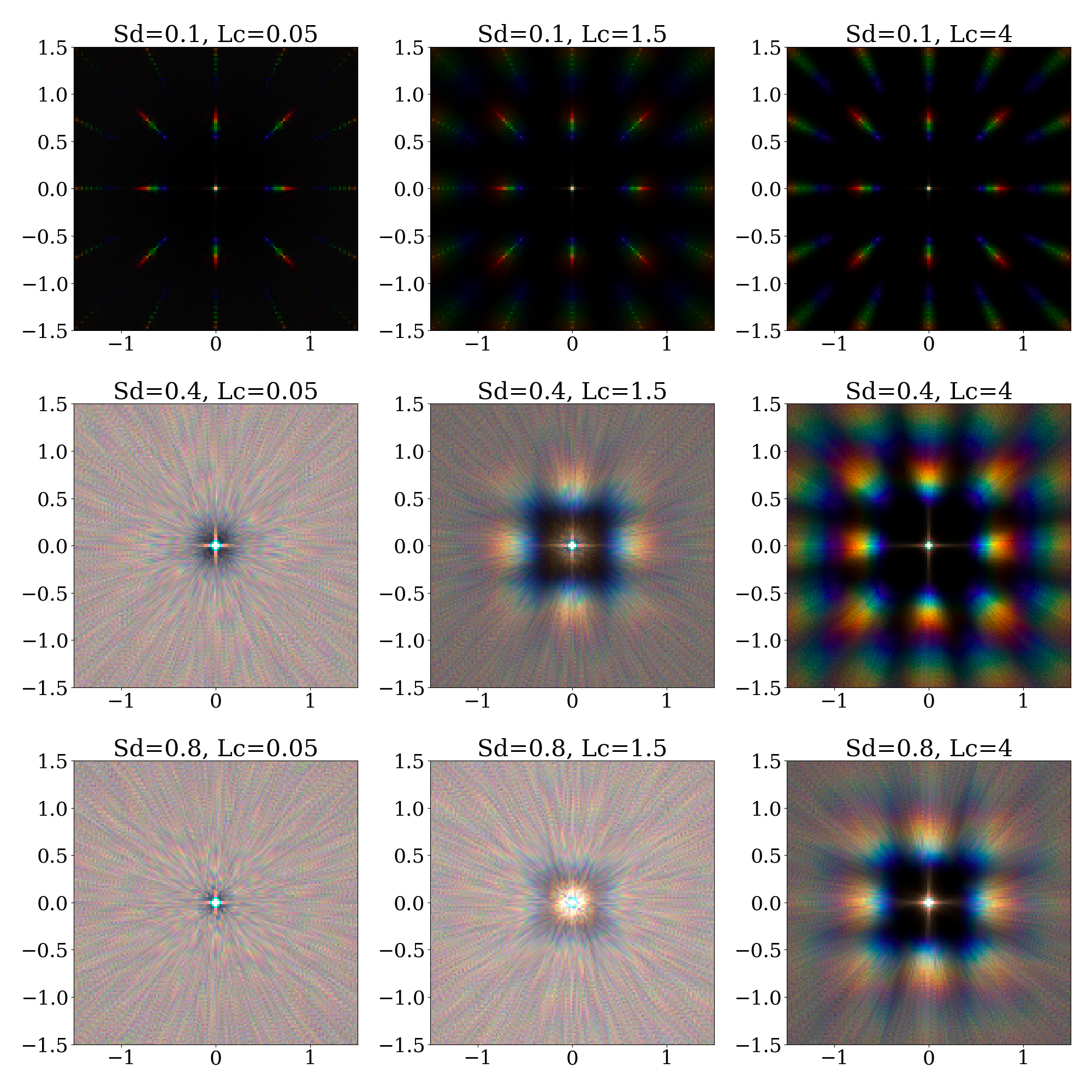}
    \caption{Scattered light patterns from disordered geometries with different degrees of disorder. The left panels show examples of 20x20 quasi-periodic ensembles of points, for different values of Sd and Lc , generated using Eq.~\ref{eq:disorder}. The right panels show calculated k-space scattering patterns obtained from incident white light by a 20 x 20 ensemble of scatterers, averaged over 5 random ensembles. Note the difference in angular properties and color clarity.}
    \label{fig:paths}
\end{figure*}

To understand how introducing disorder in a regular structure can produce such a diversity of responses, we first establish a theoretical framework for one-dimensional quasi-periodic structures with controlled perturbations.
We consider the positions $x_n$ of nanostructures defined by:
\begin{equation}
    x_n = n\times d + \epsilon_n,\label{eq:pos_def}
\end{equation}
where $d$ is the average spacing (quasi-period) and $\epsilon_n$ a random variable representing the random shift from the periodic position. We call $\epsilon$ the perturbation of the array.

We further define $\Delta_m^{(j)} = \epsilon_{j+m}-\epsilon_j$, a random variable derived from $\epsilon_n$, such that $md + \Delta_m^{(j)}$ is the distance between nanostructures $j$ and ${j+m}$. $\Delta_m^{(j)}$ thus allows a characterization and/or control of the correlations between positions at distance $m \,d$, and we call it the m-th neighbor distance noise. Controlling the $\epsilon$ and $\Delta$ random variables allows us to tune the scattering angular distribution of the structure, as will be shown below.

To control the perturbations added to the underlying periodic positions, we define
i)  the standard deviation of the random shifts $S_d = \sqrt{\mathrm{Var}(\epsilon)}$, which directly measures the intensity of the perturbation, and
ii) the correlation length $L_c$, which measures how far apart nanostructure positions are correlated. The exact definition of $L_c$ depends on the type of noise used, but it typically varies between $L_c=0$ (no correlations, \textit{i.e.,} white noise) and a few periods (up to $5\times d$). In the following, we will use $d=1$, so that $L_c$ and $S_d$ are dimensionless quantities, and our results can be scaled to any spectral range.

These definitions can be extended to 2D $(x,y)$ arrangements, by introducing independent $S_{d, x}$ and $S_{d, y}$ variables for the standard deviation of added perturbations, and $L_{c, x}$ and $L_{c, y}$ for the correlation length. Extension to arrangements where the height of each nanostructure is different and potentially correlated, \textit{i.e.} thin $(x,y,z)$ arrangements, is possible, but not discussed in this work.

Fig.~\ref{fig:paths} shows nine examples of 2D quasi-periodic arrangements (with the positions of 20 x 20 nanostructures) for different values of $S_d=S_{d, x}=S_{d, y}$ and $L_c=L_{c, x}=L_{c, y}$, along with their corresponding scattering pattern.
To generate these structures, we used Gaussian noise, with $S_d$ and $L_c$ controllable independently.
To obtain $N$ new positions (quasi-period $d$), the perturbations $\epsilon_j$ are Gaussian variables defined as :
\begin{equation}
    \epsilon_j = S_d\frac{\mathrm{Re}\left( \sum_{n=1}^{4\sigma} e^{i\phi_n} e^{-\frac{n^2}{2\sigma^2}} e^{2i\pi \frac{n\,j}{N}} \right)}{\sqrt{\sum_{n=1}^{4\sigma}  \frac{e^{-\frac{n^2}{\sigma^2}}}{2}}}\label{eq:disorder},
\end{equation}
where $\phi_n$ are independent random phases, uniformly distributed in $[0,2\pi]$, and $\sigma = \frac{N}{2\pi L_c}$. The denominator is a normalization term, equal to the standard deviation of the numerator.

With this equation, $\{\epsilon\}_j$ follows a Gaussian distribution with zero mean and variance $S_d^2$. The m-th neighbor distance noises $\{\Delta\}_m$ are also centered Gaussian variables, but with variance:
\begin{align}
    \mathrm{Var}\left(\Delta_m\right) &= 4S_d^2 \, \frac{\sum_{n=1}^{4\sigma} e^{-\frac{n^2}{\sigma^2}}  \sin\left(\pi \frac{nm}{N}\right)^2}{\sum_{n=1}^{4\sigma}  e^{-\frac{n^2}{\sigma^2}}} .
\end{align} 

In Fig.~\ref{fig:paths} the effect of $S_d$ is straightforward: it controls how far from the periodic positions each nanostructure can be, and therefore how different the scattering angular distribution is from an array of Bragg peaks.

Interpreting the effect of $L_c$, however, is more subtle: we can observe that it adds some regularity in the resulting positions and concentrates the scattered intensity around the diffraction peaks.
This is due to the fact that $L_c$ controls how many terms are involved in the summation in Eq.~\ref{eq:disorder}. For vanishing values of $L_c$, an infinite number of terms are taken into account, averaging their contributions such that $\mathrm{Var}(\epsilon_{m+j}-\epsilon_j) = \mathrm{Var}(\Delta_m) = 2S_d^2 = 2 \mathrm{Var}(\epsilon)$, corresponding to a case where the $\epsilon$ are uncorrelated, leading to a typical white noise pattern in reciprocal space (bottom left on Fig.~\ref{fig:paths}). On the other hand, a structure with a high $L_c$ will exhibit a very anisotropic scattering pattern (middle right on Fig.~\ref{fig:paths}).

To better understand how a high $L_c$ structure scatters light, we study 1D structures.


The far-field scattering of a quasi-periodic arrangement can be computed analytically, both to speed up the computations but also to better understand the underlying phenomena. To do so, we will make a few simplifying assumptions.
First, we will assume that the nanostructures emit/reflect light isotropically, to distinguish the influence of the nanostructures from that of their positions (this amounts to neglecting the form factor of the structures and only modeling the structure factor). This is particularly relevant in our case because naturally-occurring structures tend to be composed of mostly locally identical nanostructures, \textit{i.e.} the influence of their individual shape appear as a general prefactor to the scattering pattern described in the following.
Second, we will consider that the light undergoes only one scattering event, \textit{i.e.} we neglect multiple scattering paths, and that the nanostructures scatter light in a coherent manner. This is generally the case because the complete structure is typically a flat surface and only a few wavelengths large.
Lastly, because we consider a one-dimensional (1D) ensemble of nanostructures, the scattered light remains in the plane of incidence, and we can model its direction with only one variable,  the in-plane wavevector ($k_x$).

Then, the far-field scattered amplitude by our nanostructures (positions $x_n$) can be written as:
\begin{equation}
    A(k_x) = \sum_{n}e^{ik_xx_n},
\end{equation}
and the scattered intensity is given by:
\begin{align}
    I(k_x) &= AA^* = \sum_{n}e^{ik_xx_n}\sum_{m} e^{-ik_x x_m}\nonumber\\
    &= \sum_{n}\sum_{m} e^{ik_x(x_n - x_m)}.
\end{align}

Taking into account Eq.~\ref{eq:pos_def}, this leads to the following expression for the ensemble averaged scattered intensity $I$ (details in Supp Mat):
\begin{align}
        \langle I(k_x)\rangle &= N \left(1 -  |\tilde{\rho_{\epsilon}}(k_x)|^2\right) \nonumber\\
        &+ |\tilde{\rho_{\epsilon}}(k_x)|^2 \left(\frac{\sin\left(\frac{Nk_xd}{2}\right)}{\sin\left(\frac{k_xd}{2}\right)}\right)^2\label{eq:la_formule} \\
        &+\sum_{m=1}^{M} 2 (N-m)  \cos(k_xmd)\left(\tilde{\rho}_{\Delta_m}(k_x) -|\tilde{\rho_{\epsilon}}(k_x)|^2\right)\nonumber,
\end{align}
where $\tilde{\rho_\epsilon}$ (resp. $\tilde{\rho_\Delta}$) is the Fourier Transform of the probability density function of the perturbations $\epsilon$ (resp. m-th neighbor distance noises $\Delta$), and $M$ is the number of nearby neighboring positions we consider correlated (equivalently, $Md$ is the maximum distance at which we take into account correlations). Typically, $M\,d\ge 2L_c$ is sufficient with the disorder presented above.

In this formula, three terms can be identified, as highlighted by breaks in the typesetting. 
The first term corresponds to the diffuse background, typical of disordered structures, the second term is well known for Bragg peaks, as in, e.g., the grating formula, typical for periodic structures \cite{leroy2004effects}. These terms are controlled by the PDF of the perturbation, $\epsilon$, or more precisely by its Fourier Transform.
This explains the progressive disappearance of the Bragg peaks when perturbations are added, starting from the higher order modes, as $\tilde{\rho_\epsilon}$ becomes a thinner Gaussian curve when $S_d=\mathrm{Var}(\epsilon)$ increases.

\begin{figure}[ht]
    \centering
    \includegraphics[width=0.85\linewidth]{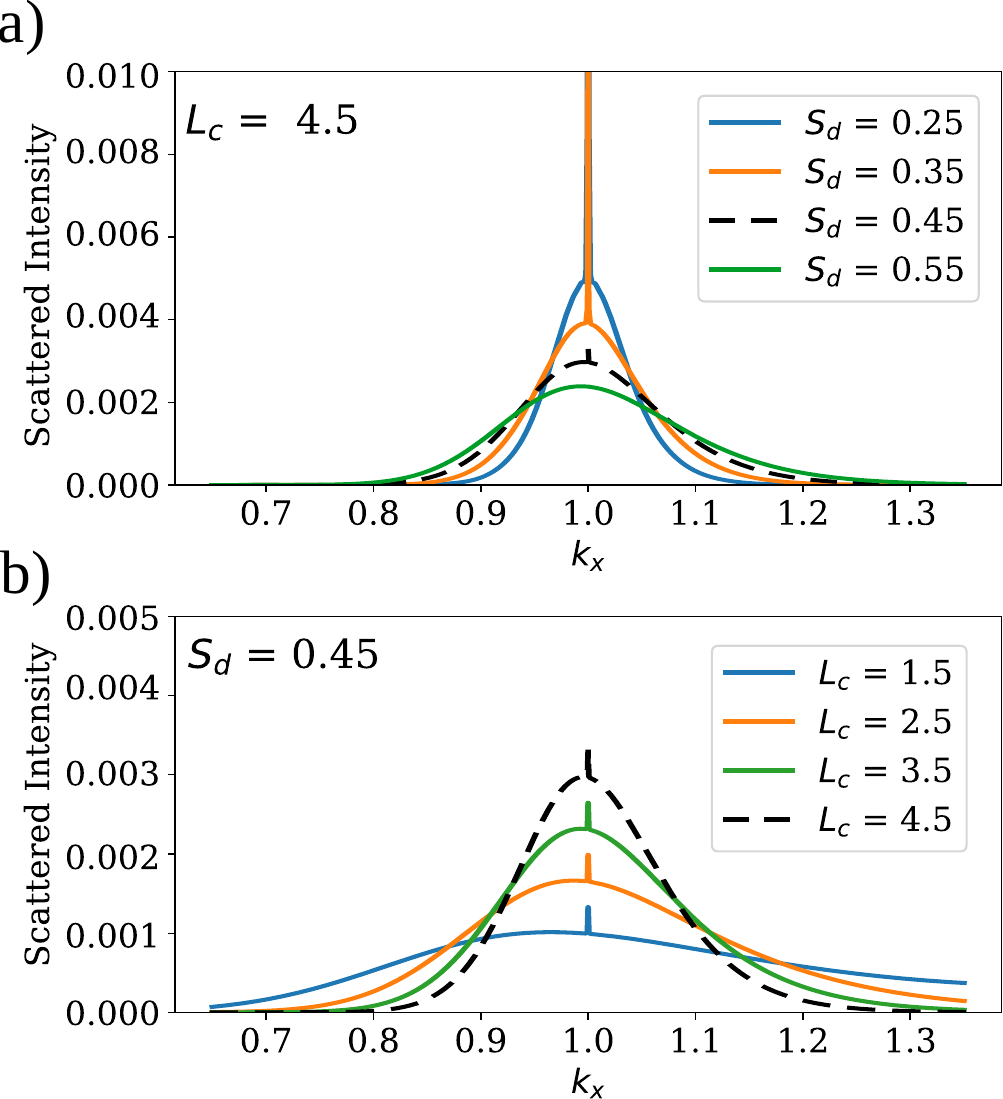}
    \caption{Evolution of the average scattering angular distribution around the first diffraction peak as a function of a) $S_d$ and b) $L_c$. The black dotted line is the one corresponding to Fig.~\ref{fig:Morpho}.}
    \label{fig:control}
\end{figure}

The third term accounts for the influence of correlations. It corresponds to a scattering component that appears in the crystallographic description of correlated disorder~\cite{welberry1985diffuse, leroy2004effects}, but has been absent from theoretical work in photonics. It is a broad component, compared to Bragg peaks, but not monotonic like the diffuse background. Its first and usually dominant contribution, in $\cos(k_x d)$, has the same periodicity as the Bragg term, so it appears around the diffraction peaks - which is why it could be mistaken for a mere broadening of the peaks. But it is not: it can produce
lobes elsewhere, it can coexist with the peaks, and, as we show below (Fig.~\ref{fig:control}), it is far more robust to noise. For these reasons, we call these features ``correlation halos''.

The halos have a further property: this term is the only one of the decomposition that can be negative. It therefore digs into the diffuse background around the specular direction, over a broad angular range. This gives a different picture of hyperuniformity than the usual one, \textit{i.e.}, a near absence of scattering near the specular direction that emerges when scatterers repel each other at short distances. Here, one can see it as the diffuse background being carved away around the specular direction by the correlation halos.

\begin{figure}[ht]
    \centering
    \includegraphics[width=0.8\linewidth]{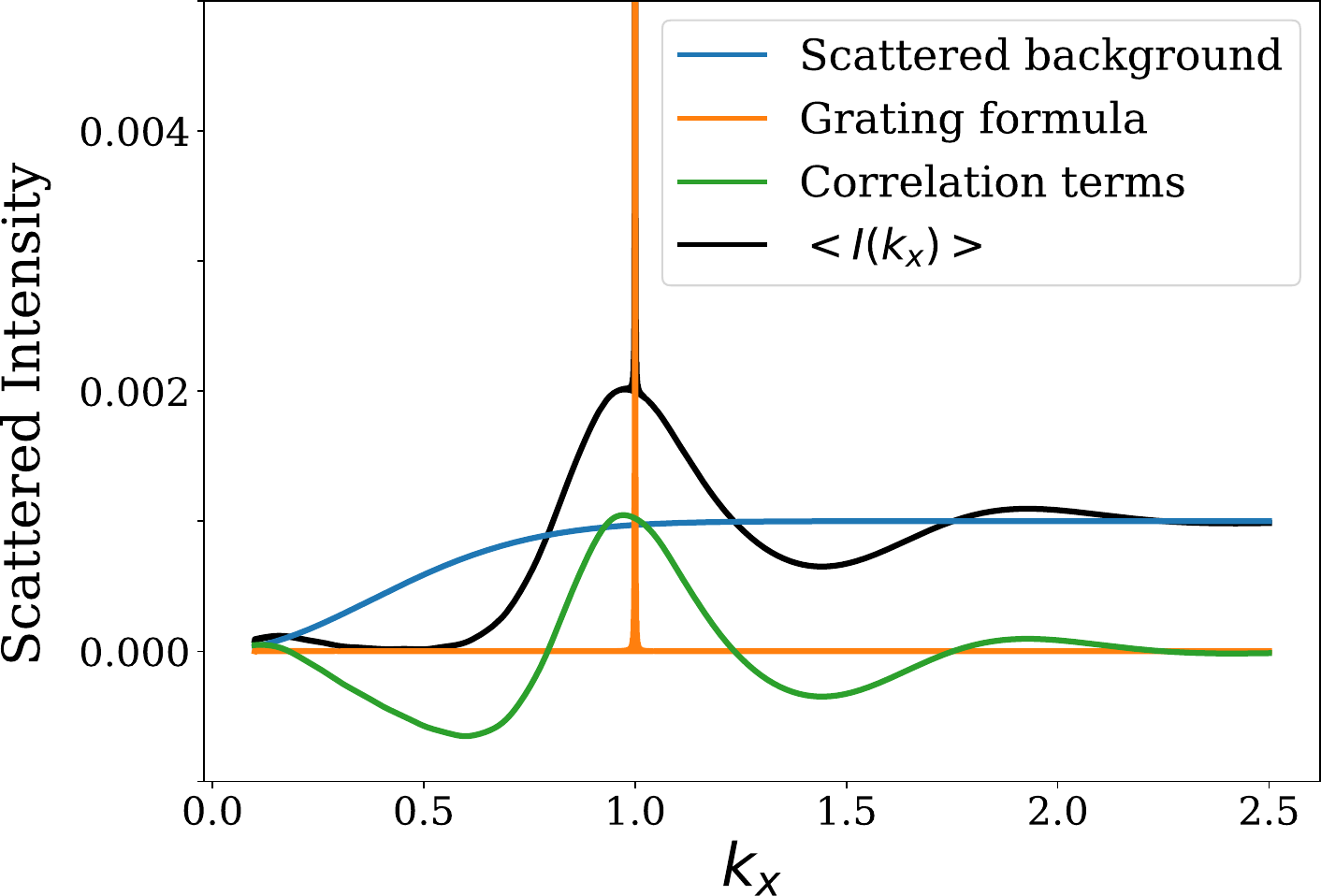}
    \caption{Average scattering angular distribution of a correlated disordered structure generated with Eq.~\ref{eq:disorder} ($S_d=0.3$, $L_c=1$), with the decomposition of the average intensity formula.}
    \label{fig:corr3}
\end{figure}

This is exactly what Fig.~\ref{fig:corr3} shows. The three components are readily identified: the diffuse background (blue) is flat and non-decreasing, the Bragg term (orange) concentrates the intensity into a sharp peak at $k_x=1$, and the correlation term (green) forms a broad, slowly varying lobe centered on that peak. On either side of the peak, the correlation term becomes negative, and this negative part plays an important role in the final scattering pattern: it is where the halo subtracts from the diffuse background. The total intensity (black) follows the flat diffuse background everywhere except there, where it is pulled below it - most clearly around $k_x \approx 0.5$. The halo does not only add a bump on the peak; it carves the background around it.

Figure~\ref{fig:control} shows a close-up of how the correlations halos are distributed around the first diffraction peak and how they are influenced by $S_d$ and $L_c$. Indeed, changing $S_d$ influences mostly the diffraction peak height, but also the angular spread of the correlation halo (see Fig.~\ref{fig:control}a). A larger $S_d$ leads to less diffraction, and for $S_d>0.5$, no diffraction peak is visible. Changing $L_c$ influences the correlation halo's angular size and maximum height, with no effect on the diffraction peak intensity (see Fig.~\ref{fig:control}b). This is expected, given that the diffraction peak intensity is only controlled by $\tilde{\rho}_\epsilon$ in Eq.~\ref{eq:la_formule}, and therefore not sensitive to $L_c$. This gives independent control of the two features: $S_d$ sets the
diffraction peak, $L_c$ sets the halo, and one can tune them separately to reach a target scattering pattern. It is also why the halos outlive the diffraction peaks: for $S_d > 0.5$ the peak has vanished, yet the halo remains — the very regime that produces the smooth, peakless patterns of natural structures such as the \textit{Morpho} wing.

\begin{figure}[ht]
    \centering
    \includegraphics[width=0.8\linewidth]{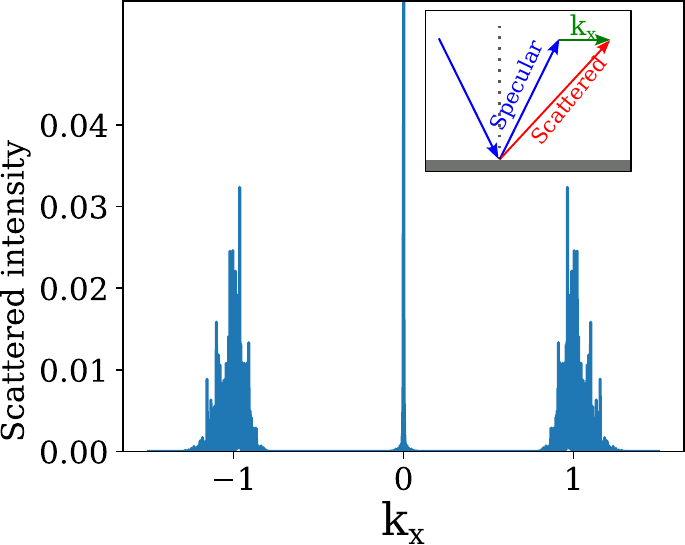}
    \caption{Scattering angular distribution by a structure with 1000 nanostructures, $S_d=0.45$, $L_c=4.5$, no averaging. (inset): Definition of the $k_x$ in-plane scattering wavevector}
    \label{fig:Morpho}
\end{figure}

We used this design space to try and reproduce one of the most well-known naturally occurring structures exhibiting structural colors, the wing scales of the Morpho butterfly, whose scattering angular distribution was measured decades ago \cite{vukusic1999quantified}. Previous works have shown that the typical blue color of the Morpho wings can be retrieved by numerical optimization tools, and that the resulting optimized structures closely resemble the Morpho wing scales \cite{barry2020evolutionary}. However, the blue structural color obtained with a periodic array of such structures is very directional, whereas the Morpho wings exhibit a smoothly varying scattering pattern centered on $\pm 60^\circ$.
To verify that correlated disorder is what gives rise to this smooth angular distribution of the scattering, we looked for a 1D quasi-periodic structure exhibiting the scattering angular distribution of the Morpho wings, with the help of a numerical optimization algorithm to find the corresponding $S_d$ and $L_c$ \cite{bennet2024illustrated}.  
Figure~\ref{fig:Morpho} shows the scattering angular distribution, as a function of $k_x$, the normalized in-plane wavevector of light, for the optimized 1D structure ($S_d=0.45$, $L_c=4.5$). We observe no diffraction peaks, but the smoothly varying lobes we have called correlation halos, here centered around the diffraction peak positions, as are typical of the Morpho wing scales, among other naturally occurring structures \cite{vukusic1999quantified, Yoshioka2004, Giraldo2016, moyroud2017disorder}. Figure~\ref{fig:control} also represents the average scattering intensity distribution for the optimized parameters (black dotted line).
This demonstrates that the correlation halos exhibited by correlated quasi-periodic structures are one of the missing blocks between the periodic and the white noise regimes.
Furthermore, it is reasonable to assume that the Morpho wing scales exhibit a strong correlation between the individual structures, given that their size is comparable to the average distance between each structure. Any shift in position of one structure will push other structures away, a behavior closely resembling what can be seen of Fig.~\ref{fig:paths}~(middle right).


\begin{figure}[ht]
    \centering
    \includegraphics[width=0.65\linewidth]{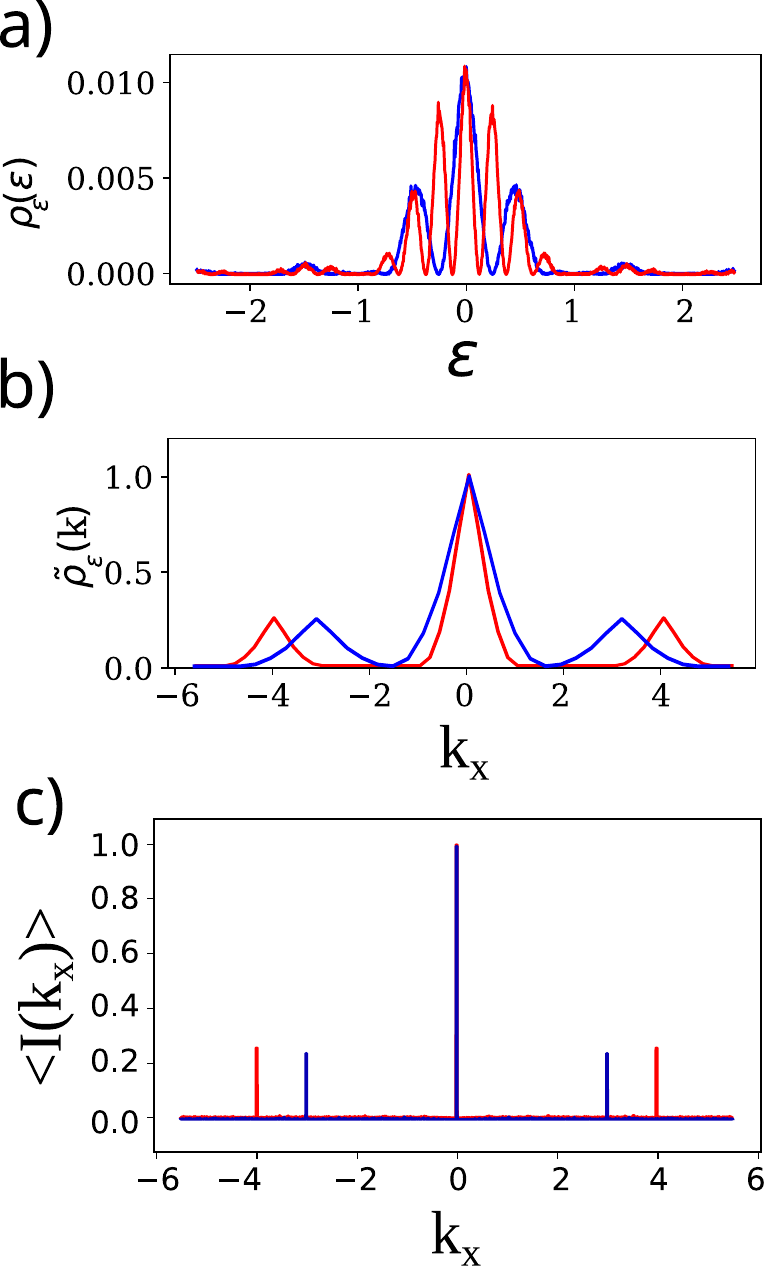}
    \caption{Example of a probability distribution selecting diffraction peaks (blue: third peak, red: fourth peak); a) $\rho(\epsilon)$ of an ensemble of nanostructures, b) Corresponding $|\tilde{\rho}_\epsilon(k)|$, c) Resulting scattering angular distribution. 1000 nanostructures, no averaging. }
    \label{fig:scat_example}
\end{figure}

Until now, we have shown how different components of the scattering angular distribution can be tuned with the $S_d$ and $L_c$ parameters, while keeping a Gaussian law for the perturbations. However, this concept can be taken further by directly engineering the probability density function of the perturbations. This way, we are able to design scattering patterns beyond Gaussian envelopes in Eq.~\ref{eq:la_formule}. In other words, even without manipulating the correlation halo term (\textit{i.e.,} without controlling $L_c$), designing specific noises (\textit{i.e.} $\epsilon$ functions), can produce tailored scattering patterns.

This is illustrated by the probability density function of Fig.~\ref{fig:scat_example}a), given by:
\begin{equation}
    \rho_{\epsilon}(\epsilon) = 2\left(\frac{\sin(\pi \epsilon)}{\pi \epsilon}\right)^2(\cos(2\pi q\epsilon)+1). \label{eq:order_select}
\end{equation}

The Fourier Transform of this function presents three triangular peaks (Fig.~\ref{fig:scat_example}b) with $q=3$ (blue) or $q=4$ (red)). Indeed, the Fourier Transform of $(\sin(x)/x)^2$ is a triangular peak centered at $k_x=0$, and the cosine term creates duplicate peaks centered around $k_x = \pm q$. As a consequence, for $q=3$ (resp. $q=4$), only diffraction peaks at $k_x=$ 0, -3 and +3 (resp. 0, -4 and +4) can be observed in the scattering pattern.
Figure~\ref{fig:scat_example}c) shows the scattering angular distribution of a structure created with this type of noise, demonstrating the extinction of all diffraction peaks except the chosen ones.

It also shows that the intensity of the remaining peaks is 0.25, with the specular reflection intensity being 1. With the approach presented here, it is impossible to reduce the energy in the specular reflection, as $\rho_{\epsilon}(\epsilon)$ must be positive. It is also impossible to design an asymmetric $I(k_x)$, as $\rho_{\epsilon}(\epsilon)$ must be real. However, in practice, the form factor of the individual nanostructures could be leveraged for additional control \cite{gralak2001morpho, boulenguez2012multiple}.

\begin{figure}
    \centering
    \includegraphics[width=0.65\linewidth]{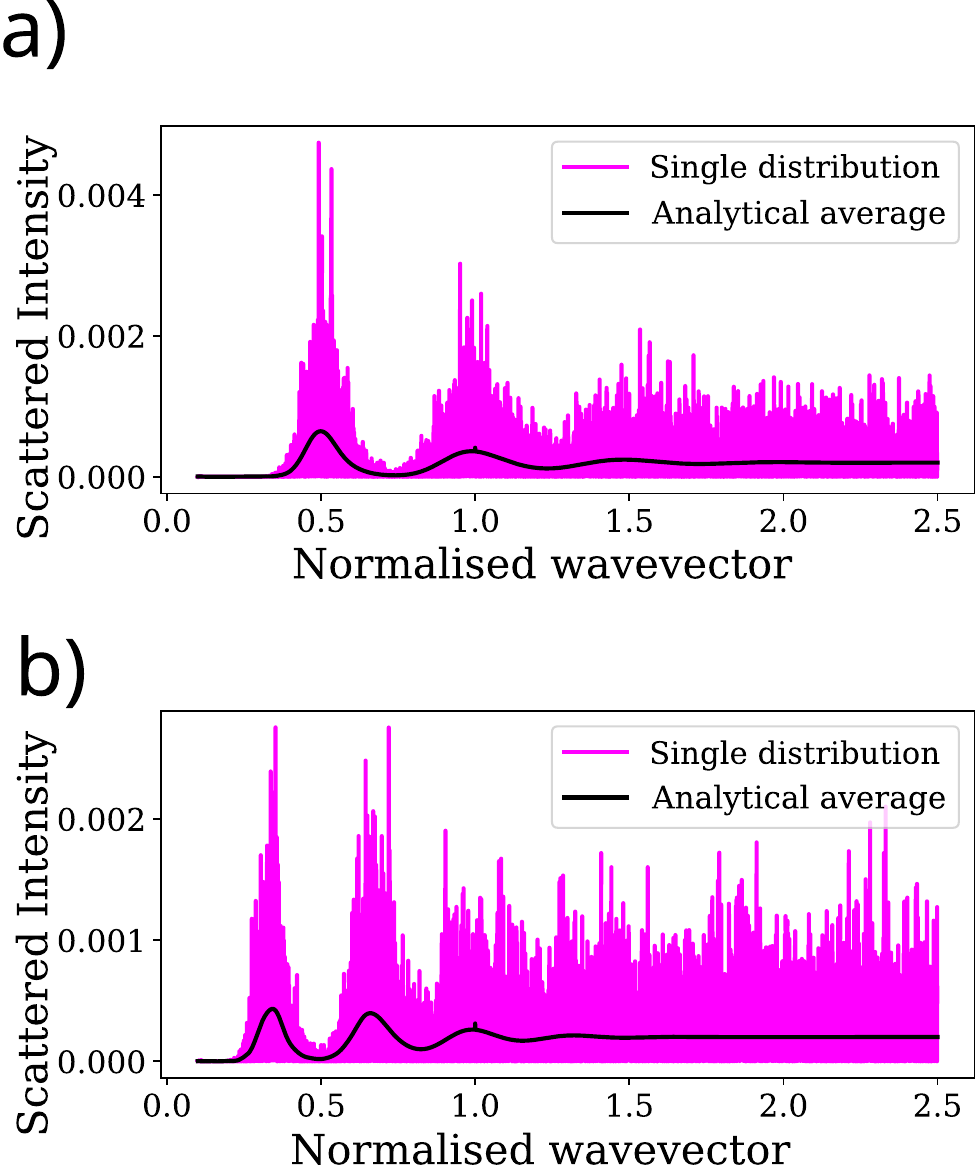}
    \caption{Scattering patterns for a handcrafted correlated disorder, where scatterers are not correlated locally, but at a distance. a) Scatterers are correlated every two positions ($c=2)$. b) Scatterers are correlated every three positions ($c=3$). Note the correlation halos appearing at other positions than Bragg peaks.}
    \label{fig:handcrafter}
\end{figure}

Similarly to the control of the PDF of $\epsilon$, it is possible to design new scattering patterns by tuning only the m-th neighbor distance noise $\Delta$. For instance, it is possible to choose specific $m$ values for which $\Delta_m$ is highly correlated, leading to different dominant $\cos(k_xmd)$ terms in Eq.\ref{eq:la_formule}. A simple way to do this is to create subgroups in the structure. One way to do this, is to define $c$, the number of subgroups, and then change the definition of the $\epsilon$ by introducing $\delta$ intermediate variables:

\begin{equation*}
    \delta_{j}  =  \sum_{i} g_{ci+j} e^{-\frac{(ci)^2}{L_c^2}},
\end{equation*}
where $g$ are Gaussian variables with zero mean and unit variance. We can then define the $\epsilon$ so they have the desired variance: $\epsilon_j = S_d\delta_j / \sqrt{Var(\delta)}$.

This lets us control the positions with correlations only within the subgroups of the structure. To give a simple example, for $c=2$, all m-th neighbor distance noises except for $m=2$ will be uncorrelated, \textit{i.e.} $\rho_{\Delta_{m}} = (\rho_{\epsilon})^2$.
In practice, it means we split between the odd and even numbered positions (\textit{e.g.}, if $j$ is even, $2i+j$ is even for all $i$), and we correlate odd positions with odd positions and even positions with even positions. One can even consider different correlation lengths for different subgroups.

Figure~\ref{fig:handcrafter} shows two examples of this, a) for $c=2$, where a correlation halo appears at $k=0.5$, and b) for $c=3$, where correlation halos appear at $k=1/3$ and $k=2/3$.
This paves the way for a very versatile scattering pattern design space.

All these properties generalize well to two dimensional (2D) structures, where the nanostructure positions are $\{(x,y)\}_n$. The derivation of the disorder generation and the scattered intensity formulas are found in the Supp. Mat.

To conclude, more than a model for a given structure, our work proposes a design framework. Its power has been demonstrated on the Morpho wing, retrieving the noise amplitude and the degree of correlation that produce a correlation halo in the absence of any diffraction peak. To go further, because the scattering is written explicitly in terms of the noise statistics, one can search for the probability density functions — of the perturbations or of the m-th neighbor distances — that produce a target scattering pattern, and so begin to design scattering itself rather than merely characterize it. The individual form factor of the scatterers, deliberately set aside here, provides a further tool — one that lifts the constraints a real, positive PDF imposes on the achievable scattering patterns.

More broadly, our results show that there is no single route from order to disorder. Perturbation statistics, correlations and m-th neighbor distances each open a distinct path, with its own optical signature — halos that persist where Bragg peaks have vanished, peaks selectively switched off, features displaced away from the Bragg positions. Even the case of "hyperuniformity" includes structures whose scattering can differ strongly, and our design framework provides a direct link between the type of correlation imposed on the disorder and the resulting suppression of small-angle scattering. 
It thus offers not only a way to map the different regimes of hyperuniformity, but also design tools to engineer them. Diffuse scattering, long treated as an unavoidable whitish background, thus becomes a designable quantity, expanding the range of available structures for structural color and metasurfaces.


\bibliography{article}

@Article{Yu2021engineered,
  author    = {Yu, Sunkyu and Qiu, Cheng-Wei and Chong, Yidong and Torquato, Salvatore and Park, Namkyoo},
  journal   = {Nature Reviews Materials},
  title     = {Engineered disorder in photonics},
  year      = {2021},
  number    = {3},
  pages     = {226--243},
  volume    = {6},
  publisher = {Nature Publishing Group},
}

@article{torquato2003local,
  title={Local density fluctuations, hyperuniformity, and order metrics},
  author={Torquato, Salvatore and Stillinger, Frank H},
  journal={Physical Review E},
  volume={68},
  number={4},
  pages={041113},
  year={2003},
  publisher={APS}
}

@article{sterl2021shaping,
  title={Shaping the color and angular appearance of plasmonic metasurfaces with tailored disorder},
  author={Sterl, Florian and Herkert, Ediz and Both, Steffen and Weiss, Thomas and Giessen, Harald},
  journal={ACS Nano},
  volume={15},
  number={6},
  pages={10318--10327},
  year={2021},
  publisher={ACS Publications}
}

@article{herkert2023influence,
  title={Influence of structural disorder on plasmonic metasurfaces and their colors—a coupled point dipole approach: tutorial},
  author={Herkert, Ediz and Sterl, Florian and Both, Steffen and Tikhodeev, Sergei G and Weiss, Thomas and Giessen, Harald},
  journal={JOSA B},
  volume={40},
  number={3},
  pages={B59--B99},
  year={2023},
  publisher={Optica Publishing Group}
}

@article{vynck2022visual,
  title={The visual appearances of disordered optical metasurfaces},
  author={Vynck, Kevin and Pacanowski, Romain and Agreda, Adrian and Dufay, Arthur and Granier, Xavier and Lalanne, Philippe},
  journal={Nature Materials},
  volume={21},
  number={9},
  pages={1035--1041},
  year={2022},
  publisher={Nature Publishing Group UK London}
}

@article{lalanne2024disordered,
  title={Disordered optical metasurfaces: basics, design and applications},
  author={Lalanne, PHILIPPE and Chen, MIAO and Rockstuhl, CARSTEN and Sprafke, ALEXANDER and Dmitriev, ALEXANDRE and Vynck, KEVIN},
  journal={arXiv preprint arXiv:2408.09425},
  year={2024}
}

@article{gralak2001morpho,
  title={\textit{Morpho} butterflies wings color modeled with lamellar grating theory},
  author={Gralak, Boris and Tayeb, G{\'e}rard and Enoch, Stefan},
  journal={Optics express},
  volume={9},
  number={11},
  pages={567--578},
  year={2001},
  publisher={Optica Publishing Group}
}

@article{boulenguez2012multiple,
  title={Multiple scaled disorder in the photonic structure of \textit{Morpho} rhetenor butterfly},
  author={Boulenguez, J and Berthier, Serge and Leroy, F},
  journal={Applied Physics A},
  volume={106},
  pages={1005--1011},
  year={2012},
  publisher={Springer}
}

@article{vukusic1999quantified,
  title={Quantified interference and diffraction in single \textit{Morpho} butterfly scales},
  author={Vukusic, P and Sambles, JR and Lawrence, CR and Wootton, RJ},
  journal={Proceedings of the Royal Society of London. Series B: Biological Sciences},
  volume={266},
  number={1427},
  pages={1403--1411},
  year={1999},
  publisher={The Royal Society}
}

@article{saito2011numerical,
  title={Numerical analysis on the optical role of nano-randomness on the \textit{Morpho} butterfly's scale},
  author={Saito, Akira and Yonezawa, Masaru and Murase, Junichi and Juodkazis, Saulius and Mizeikis, Vygantas and Akai-Kasaya, Megumi and Kuwahara, Yuji},
  journal={Journal of Nanoscience and Nanotechnology},
  volume={11},
  number={4},
  pages={2785--2792},
  year={2011},
  publisher={American Scientific Publishers}
}

@article{schroeder2018s,
  title={It's not a bug, it's a feature: functional materials in insects},
  author={Schroeder, Thomas BH and Houghtaling, Jared and Wilts, Bodo D and Mayer, Michael},
  journal={Advanced Materials},
  volume={30},
  number={19},
  pages={1705322},
  year={2018},
  publisher={Wiley Online Library}
}

@article{song2017reproducing,
  title={Reproducing the hierarchy of disorder for \textit{Morpho}-inspired, broad-angle color reflection},
  author={Song, Bokwang and Johansen, Villads Egede and Sigmund, Ole and Shin, Jung H},
  journal={Scientific reports},
  volume={7},
  number={1},
  pages={46023},
  year={2017},
  publisher={Nature Publishing Group UK London}
}

@article{mouchet2020optical,
  title={Optical costs and benefits of disorder in biological photonic crystals},
  author={Mouchet, S{\'e}bastien R and Luke, Stephen and McDonald, Luke T and Vukusic, Pete},
  journal={Faraday Discussions},
  volume={223},
  pages={9--48},
  year={2020},
  publisher={Royal Society of Chemistry}
}

@article{johansen2014optical,
  title={Optical role of randomness for structured surfaces},
  author={Johansen, Villads Egede},
  journal={Applied Optics},
  volume={53},
  number={11},
  pages={2405--2415},
  year={2014},
  publisher={Optica Publishing Group}
}

@article{wilts2020ultra,
  title={Ultra-dense, curved, grating optics determines peacock spider coloration},
  author={Wilts, Bodo D and Otto, J{\"u}rgen and Stavenga, Doekele G},
  journal={Nanoscale advances},
  volume={2},
  number={3},
  pages={1122--1127},
  year={2020},
  publisher={Royal Society of Chemistry}
}

@article{moyroud2017disorder,
  title={Disorder in convergent floral nanostructures enhances signalling to bees},
  author={Moyroud, Edwige and Wenzel, Tobias and Middleton, Rox and Rudall, Paula J and Banks, Hannah and Reed, Alison and Mellers, Greg and Killoran, Patrick and Westwood, M Murphy and Steiner, Ullrich and others},
  journal={Nature},
  volume={550},
  number={7677},
  pages={469--474},
  year={2017},
  publisher={Nature Publishing Group UK London}
}

@article{leroy2004effects,
  title={Effects of near-neighbor correlations on the diffuse scattering from a one-dimensional paracrystal},
  author={Leroy, Fr{\'e}d{\'e}ric and Lazzari, R{\'e}mi and Renaud, Gilles},
  journal={Acta Crystallographica Section A: Foundations of Crystallography},
  volume={60},
  number={6},
  pages={565--581},
  year={2004},
  publisher={International Union of Crystallography}
}

@article{vynck2023light,
  title={Light in correlated disordered media},
  author={Vynck, Kevin and Pierrat, Romain and Carminati, R{\'e}mi and Froufe-P{\'e}rez, Luis S and Scheffold, Frank and Sapienza, Riccardo and Vignolini, Silvia and S{\'a}enz, Juan Jos{\'e}},
  journal={Reviews of Modern Physics},
  volume={95},
  number={4},
  pages={045003},
  year={2023},
  publisher={APS}
}

@article{leseur2016high,
  title={High-density hyperuniform materials can be transparent},
  author={Leseur, Olivier and Pierrat, Romain and Carminati, R{\'e}mi},
  journal={Optica},
  volume={3},
  number={7},
  pages={763--767},
  year={2016},
  publisher={Optica Publishing Group}
}

@article{bigourdan2019enhanced,
  title={Enhanced absorption of waves in stealth hyperuniform disordered media},
  author={Bigourdan, Florian and Pierrat, Romain and Carminati, R{\'e}mi},
  journal={Optics Express},
  volume={27},
  number={6},
  pages={8666--8682},
  year={2019},
  publisher={Optica Publishing Group}
}

@article{rothammer2021tailored,
  title={Tailored disorder in photonics: learning from nature},
  author={Rothammer, Maximilian and Zollfrank, Cordt and Busch, Kurt and von Freymann, Georg},
  journal={Advanced Optical Materials},
  volume={9},
  number={19},
  pages={2100787},
  year={2021},
  publisher={Wiley Online Library}
}

@article{Klatt_phoamtonics,
author = {Michael A. Klatt  and Paul J. Steinhardt  and Salvatore Torquato },
title = {Phoamtonic designs yield sizeable 3D photonic band gaps},
journal = {Proceedings of the National Academy of Sciences},
volume = {116},
number = {47},
pages = {23480-23486},
year = {2019},
doi = {10.1073/pnas.1912730116}}

@article{Jacucci_corrdis,
author = {Gianni Jacucci  and Silvia Vignolini  and Lukas Schertel },
title = {The limitations of extending nature’s color palette in correlated, disordered systems},
journal = {Proceedings of the National Academy of Sciences},
volume = {117},
number = {38},
pages = {23345-23349},
year = {2020},
doi = {10.1073/pnas.2010486117}}

@article{StavengaKingfisher,
    author = {Stavenga, Doekele G. and Tinbergen, Jan and Leertouwer, Hein L. and Wilts, Bodo D.},
    title = {Kingfisher feathers – colouration by pigments, spongy nanostructures and thin films},
    journal = {Journal of Experimental Biology},
    volume = {214},
    number = {23},
    pages = {3960-3967},
    year = {2011},
    month = {12},
    issn = {0022-0949},
    doi = {10.1242/jeb.062620}
}

@article{DjeghdiAnt,
author = {Djeghdi, Kenza and Steiner, Ullrich and Wilts, Bodo D.},
title = {3D Tomographic Analysis of the Order-Disorder Interplay in the \textit{Pachyrhynchus congestus mirabilis} Weevil},
journal = {Advanced Science},
volume = {9},
number = {26},
pages = {2202145},
year = {2022}
}

@article{YinMacaw,
author = {Haiwei Yin  and Biqin Dong  and Xiaohan Liu  and Tianrong Zhan  and Lei Shi  and Jian Zi  and Eli Yablonovitch },
title = {Amorphous diamond-structured photonic crystal in the feather barbs of the scarlet macaw},
journal = {Proceedings of the National Academy of Sciences},
volume = {109},
number = {27},
pages = {10798-10801},
year = {2012},
doi = {10.1073/pnas.1204383109}
}

@article{Giraldo2016,
    author = {Giraldo, M. A. and Yoshioka, S. and Liu, C. and Stavenga, D. G.},
    title = {Coloration mechanisms and phylogeny of \textit{Morpho} butterflies},
    journal = {Journal of Experimental Biology},
    volume = {219},
    number = {24},
    pages = {3936-3944},
    year = {2016},
    month = {12},
    issn = {0022-0949},
    doi = {10.1242/jeb.148726}
}

@article{Yoshioka2004,
    author = {Yoshioka, Shinya and Kinoshita, Shuichi},
    title = {Wavelength–selective and anisotropic light–diffusing scale on the wing of the \textit{Morpho} butterfly},
    journal = {Proceedings of the Royal Society B: Biological Sciences},
    volume = {271},
    number = {1539},
    pages = {581-587},
    year = {2004},
    month = {03},
    issn = {0962-8452},
    doi = {10.1098/rspb.2003.2618}
}

@article{welberry1985diffuse,
  title={Diffuse X-ray scattering and models of disorder},
  author={Welberry, TR},
  journal={Reports on Progress in Physics},
  volume={48},
  number={11},
  pages={1543},
  year={1985},
  publisher={IOP Publishing}
}

@article{chen2025emergent,
  title={Emergent scattering regimes in disordered metasurfaces near critical packing},
  author={Chen, Miao and Agreda, Adrian and Wu, Tong and Carcenac, Franck and Vynck, Kevin and Lalanne, Philippe},
  journal={Nature Communications},
  volume={16},
  number={1},
  pages={11125},
  year={2025},
  publisher={Nature Publishing Group UK London}
}

@article{barry2020evolutionary,
  title={Evolutionary algorithms converge towards evolved biological photonic structures},
  author={Barry, Mamadou Aliou and Berthier, Vincent and Wilts, Bodo D and Cambourieux, Marie-Claire and Bennet, Pauline and Poll{\`e}s, R{\'e}mi and Teytaud, Olivier and Centeno, Emmanuel and Biais, Nicolas and Moreau, Antoine},
  journal={Scientific reports},
  volume={10},
  number={1},
  pages={12024},
  year={2020},
  publisher={Nature Publishing Group UK London}
}

@article{bennet2024illustrated,
  title={Illustrated tutorial on global optimization<? TeX$\backslash$break?> in nanophotonics},
  author={Bennet, Pauline and Langevin, Denis and Essoual, Chaymae and Khaireh-Walieh, Abdourahman and Teytaud, Olivier and Wiecha, Peter and Moreau, Antoine},
  journal={Journal of the Optical Society of America B},
  volume={41},
  number={2},
  pages={A126--A145},
  year={2024},
  publisher={Optica Publishing Group}
}

\end{document}